\documentclass[jmp,amsmath,amssymb,preprint]{revtex4-1}
\usepackage{dcolumn}
\usepackage{bm}
\usepackage{color}
\usepackage{rotating}
\usepackage{graphicx}
\usepackage{latexsym}
\usepackage{amssymb}
\usepackage{amsfonts}
\usepackage{soul}
\usepackage{color}
\usepackage{amsmath}

\begin{document}


\title{Strongly anisotropic spin relaxation in graphene/transition metal dichalcogenide heterostructures at room temperature}

\author{L. Antonio Ben\'{i}tez$^{1,2}$}
\email{antonio.benitez@icn2.cat}
\author{Juan F. Sierra$^{1}$}
\thanks{These authors contributed equally to this work}
\author{Williams Savero Torres$^{1}$}
\thanks{These authors contributed equally to this work}
\author{Alo\"{i}s Arrighi$^{1,2}$}
\author{Fr\'{e}d\'{e}ric Bonell$^{1}$}
\author{Marius V. Costache$^1$}
\author{Sergio O. Valenzuela$^{1,3}$}
\email{SOV@icrea.cat}
\affiliation{$^1$Catalan Institute of Nanoscience and Nanotechnology (ICN2), CSIC and The Barcelona Institute of Science and Technology (BIST), Campus UAB, Bellaterra, 08193 Barcelona, Spain}
\affiliation{$^2$Universitat Auton\'oma de Barcelona, Bellaterra, 08193 Barcelona, Spain}
\affiliation{$^3$Instituci\'{o} Catalana de Recerca i Estudis Avan\c{c}ats (ICREA), 08070 Barcelona, Spain}



\begin{abstract}
\textbf{A large enhancement in the spin-orbit coupling of graphene has been predicted when interfacing it with semiconducting transition metal dichalcogenides. Signatures of such an enhancement have been reported but the nature of the spin relaxation in these systems remains unknown. Here, we unambiguously demonstrate anisotropic spin dynamics in bilayer heterostructures comprising graphene and tungsten or molybdenum disulphide (WS$_{2}$, MoS$_{2}$). We observe that the spin lifetime varies over one order of magnitude depending on the spin orientation, being largest when the spins point out of the graphene plane. This indicates that the strong spin-valley coupling in the transition metal dichalcogenide is imprinted in the bilayer and felt by the propagating spins. 
These findings provide a rich platform to explore coupled spin-valley phenomena and offer novel spin manipulation strategies based on spin relaxation anisotropy in two-dimensional materials.}

\end{abstract}

\maketitle

Graphene has emerged as the foremost material for future two-dimensional spintronics due to its tuneable electronic properties \cite{han2014,SR2014,SR2015,SOV2017}. Spin information can be transported over long distances \cite{kamalakar2015,drogeler2016} and, in principle, be manipulated by using  magnetic correlations or large spin-orbit coupling (SOC) induced by proximity effects \cite{SOV2017,marchenko2012,klimovskikh2014,avsar2014,gmitra2015,gmitra2016,wang2015,wang2016,yang2016,yang2017,vaklinova2016,dushenko2016,WST2017,garcia2017,offidani2017}. First principle calculations have shown that semiconducting transition metal dichalcogenides (TMDC) can induce a SOC in graphene in the meV range while preserving its linear Dirac band structure \cite{gmitra2015, gmitra2016}. The enhancement of the SOC has been demonstrated using non-local charge transport and weak (anti-)localization measurements \cite{avsar2014,wang2015,wang2016,yang2016,yang2017}, the most salient signature being a drastic reduction of the spin lifetime down to the picosecond range. In recent spin transport experiments, the spin sink effect in TMDCs was used to modulate the transmission of pure spin currents. This enabled the realization of a spin field-effect switch that changes between ``on" and ``off" by gate tuning \cite{yan2016,dankert2017}.

A further consequence of the proximity-induced SOC by a TMDC, which has not been addressed experimentally, is the imprint of the strong spin-valley coupling \cite{wang2012,xiao2012,yang2015,ye2016,rivera2016}.  As depicted in Fig. 1a, a band gap should open in the graphene Dirac cone due to the breaking of pseudospin symmetry, while the SOC, combined with broken space inversion symmetry, removes the spin degeneracy \cite{gmitra2015, gmitra2016}. The spins in these bands tilt out of the graphene plane, with the out-of-plane component alternating from up to down as the energy of the bands increases, in a sequence that inverts between the $K$ and $K'$ points of the Brillouin zone.

The spin splitting and texture in graphene/TMDC imply that the spin dynamics would likely differ for spins pointing in ($\parallel$) and out of ($\perp$) the graphene plane, leading to distinct spin lifetimes, $\tau_\mathrm{s}^\parallel$ and $\tau_\mathrm{s}^\perp$, and associated spin relaxation lengths, $\lambda_\mathrm{s}^\parallel$ and $\lambda_\mathrm{s}^\perp$. Indeed, realistic modelling predicts that the spin lifetime anisotropy ratio $\zeta\equiv\tau_\mathrm{s}^\perp/\tau_\mathrm{s}^\parallel=(\lambda_\mathrm{s}^\perp/\lambda_\mathrm{s}^\parallel)^2$ could reach values as large as a few hundreds, in the presence of intervalley scattering \cite{cummings2017}. Quantification of $\zeta$ can therefore provide unique insight into spin-valley coupling mechanisms and help elucidate the nature of the induced SOC in graphene \cite{han2014,BR2016}.

Recent experiments have demonstrated that $\zeta$ can be determined combining in-plane and out-of plane spin precession measurements \cite{BR2016,BR2017}. In order to reveal the nature of the spin dynamics in graphene/TMDC, we implement such a technique using the device depicted in Fig. 1b, with TMDC being either WS$_{2}$ or MoS$_{2}$. The experiments are based on the standard non-local spin injection and detection approach \cite{js1985,jedema2001,SOV2009}.
A multilayer TMDC flake is placed over graphene between the ferromagnetic injector (F1) and detector (F2) electrodes, creating a graphene/TMDC van der Walls heterostructure (see Methods and Fig. 1c). The TMDC modifies the graphene band structure by proximity effect and, as a consequence, the spin relaxation. Considering that $\tau_\mathrm{s}^\parallel$ in the modified graphene region is expected to be in the range of a few ps \cite{dankert2017,cummings2017}, the spin relaxation length should be in the submicron range. The width $w$ of the TMDC flake is thus selected to be about one micrometre so as not to completely suppress the spin population when spins are in-plane. The spin channel length $L$, which is defined as the distance between F1 and F2, is much longer than $w$ (about 10 $\mu$m) to ensure that the spin precession can be studied at moderate magnetic fields \cite{BR2016,BR2017}. A back-gate voltage is applied to the substrate to tune the spin absorption in the TMDC \cite{yan2016,dankert2017} while all the measurements are acquired at room temperature.

Owing to magnetic shape anisotropy, the magnetizations of F1 and F2 tend to be in-plane. A perpendicular magnetic field $B$ causes spins to precess exclusively in plane and senses $\tau_\mathrm{s}^\parallel$ only. In order to obtain $\tau_\mathrm{s}^\perp$, the strategies represented in Figs. 1d and 1e are followed. These strategies are described in Refs. \cite{BR2016,BR2017} and rely on the application of an oblique $B$ characterized by an angle $\beta$ (Fig. 1d) or an in-plane $B$ perpendicular to the easy magnetization axes of F1/F2 (Fig. 1e). As represented by the red arrows, such magnetic fields force the spins to precess out of the plane as they diffuse towards the detector. When a spin reaches the graphene/TMDC region, its orientation relative to the graphene plane is characterized by the angle $\beta^*$, which depends on the magnitude of $B$ (Fig. 1b). The spin precession dynamics therefore becomes sensitive to both $\tau_\mathrm{s}^\parallel$ and $\tau_\mathrm{s}^\perp$, and $\tau_\mathrm{s}^\perp$ can be determined.

Figure 2 demonstrates the changes in the spin precession lineshape, $R_\mathrm{nl}$ vs $B$, induced by the TMDC. The non-local spin resistance $R_\mathrm{nl}=V_\mathrm{nl}/{I}$ is determined from the voltage $V_\mathrm{nl}$ at the detector F2, which is generated by a current $I$ flowing at the injector F1 (Fig. 1b). Figures 2a to 2c show measurements for a typical graphene/WS$_{2}$ device (Device 1). The data are acquired for parallel and antiparallel configurations of the F1/F2 magnetizations. Figures 2d to 2f show equivalent measurements for a reference device (without WS$_{2}$), which was fabricated in the same graphene flake (Fig. 1c).

The spin precession response in the two devices is strikingly different. Figures 2a and 2d present conventional spin precession measurements with an out-of-plane magnetic field. Even though spin precession is observed in both cases, $|R_\mathrm{nl}|$ in the graphene/WS$_{2}$ device (Fig. 2a) is two orders of magnitude smaller than in the reference (Fig. 2d).
The decrease in $|R_\mathrm{nl}|$ indicates a large reduction of $\lambda_\mathrm{s}^\parallel$ in the graphene/WS$_{2}$ region, as observed previously for MoS$_2$ \cite{dankert2017}. The value of $\lambda_\mathrm{s}^\parallel$ can be determined by solving the diffusive equations at $B=0$ (Supplementary Information). From the change in $R_\mathrm{nl,0} = R_\mathrm{nl}(B=0)$ between the graphene/WS$_2$ and the reference devices, $\lambda_\mathrm{s}^\parallel$ is estimated to be about 0.2-0.4 $\mu$m, which is significantly smaller than the typical $\lambda_\mathrm{s,gr} \sim$ 3-5 $\mu$m in our pristine graphene.

Figures 2b and 2e present spin precession measurements for in-plane $B$, as shown in Fig. 1e. While the change in the $B$ orientation results in no significant variation in the reference device, the changes observed in the graphene/WS$_{2}$ device are remarkable (compare Figs. 2a and 2b). Figure 2b shows that, as $|B|$ increases, $|R_\mathrm{nl}|$ becomes much larger than its value at $B=0$. The anomalous enhancement of $|R_\mathrm{nl}|$ is a clear indication of anisotropic spin relaxation, with $\lambda_\mathrm{s}^\perp$ larger than $\lambda_\mathrm{s}^\parallel$ and thus $\zeta > 1$. Similar results for three other devices are shown in Supplementary Fig. 1, one of them with MoS$_2$, which demonstrates that the anisotropic relaxation is not limited to graphene/WS$_2$.

The difference between Figs. 2d and 2e is due to the tilting of the F1/F2 magnetizations with $B$. The tilting angle $\gamma$, which is calculated from the fittings to the spin precession in the reference \cite{BR2016}, is more pronounced for in-plane $B$, for which the shape anisotropy is smaller. In an isotropic system, the non-local resistance has the form $R_\mathrm{nl,iso}^{\pm}=[\pm g(B)\cos^2 \gamma + \sin^2 \gamma]R_\mathrm{nl,0}$ for initially parallel (+) and antiparallel (-) magnetization configurations, with $g(B)$ a function that captures the precession response \cite{BR2017,jedema2001,SOV2009}. By defining $\Delta R_\mathrm{nl}= R_\mathrm{nl}^{+}-R_\mathrm{nl}^{-}$, it is evident that, for an isotropic system and small $\gamma$, $\Delta R_\mathrm{nl} \approx 2 g(B)R_\mathrm{nl,0}$ is independent of the $B$ orientation. The obtained $\Delta R_\mathrm{nl}$ for $B$ in-plane (red) and out-of-plane (blue) are shown in Figs. 2c and 2f for the graphene/WS$_2$ and the reference device, respectively. The nearly perfect overlap of the two curves in Fig. 2f is a consequence of the isotropic spin relaxation in graphene \cite{BR2016,BR2017} while the disparity of the curves in Fig. 2c further demonstrates the highly anisotropic nature of the spin transport in graphene/WS$_{2}$. The extrema in $R_\mathrm{nl}$ are reached when the aggregate orientation of the diffusing spins have rotated by $\sim\pi/2$ at the WS$_{2}$ location. Because the diffusing spins pass by WS$_{2}$ before F2, this occurs at magnetic fields that are slightly larger than those at which $R_\mathrm{nl}=0$ in the conventional spin precession measurements (dashed vertical lines in Fig. 2c).

Having determined $\lambda_\mathrm{s}^\parallel$, $\zeta$ can be obtained using out-of-plane precession with oblique $B$. Because the spins diffuse towards WS$_2$, they are characterized by a broad distribution of inclination angles $\beta^*$ (Fig. 1b). As $B$ increases, the spin component that is perpendicular to the magnetic field dephases and only the component parallel to $B$ contributes to $R_\mathrm{nl}$ \cite{BR2016}. In the case depicted in Fig. 1d, this component is non-zero and the spin orientation is univocally determined by $\beta^*=\beta$, which greatly simplifies the analysis to obtain $\lambda_\mathrm{s}^\perp$ and $\zeta$. Indeed, the effective spin relaxation length in graphene/WS$_2$ for arbitrary $\beta$, $\lambda^{\beta}_\mathrm{s}$, can be calculated using the same procedure that is used to find $\lambda_\mathrm{s}^\parallel$ (Supplementary Information). Note that, to achieve full dephasing at moderate $B$, the WS$_{2}$ flake is close to the ferromagnetic electrode that typically plays the role of detector (F2 in Fig. 1b).

Figure 3a shows spin precession measurements for a representative set of $\beta$ values for Device 2, with F2 as detector. A back-gate voltage $V_\mathrm{g} = -15$ V is applied to suppress the spin absorption in WS$_2$ (see below). It is observed that diffusive broadening dephases the precessional motion at $B_\mathrm{d} \sim 0.12$ T. For $B>B_\mathrm{d}$, $R_\mathrm{nl}$ is nearly constant with increasing $B$, with a magnitude $R_\mathrm{nl}^{\beta}\equiv R_\mathrm{nl} (B>B_\mathrm{d})$. At $\beta=90^{\circ}$, the response is similar to that in Fig. 2a. However, as soon as $B$ is tilted from the perpendicular orientation, $R_\mathrm{nl}^{\beta}$ increases anomalously, and a few degrees tilt results in $R_\mathrm{nl}^{\beta}>R_\mathrm{{nl,0}}$ even at small $B$ (e.g. for $\beta=85.5^{\circ}$). This is in stark contrast to the case of pristine graphene. Equivalent measurements for a reference device are shown in Supplementary Fig. 2; there $\zeta \approx 1$ and $R_\mathrm{{nl,0}}$ is an upper limit for $R_\mathrm{nl}$ \cite{BR2016}. As a comparison, Fig. 3b shows measurements with the role of F1 and F2 reversed. Because the injector (F2) is now close to WS$_2$, the aggregate spin precession angle at the WS$_2$ location for any given $B$, is smaller than in the case where the injector is far away (Fig. 3a). Full dephasing at graphene/WS$_2$ is not achieved, while $R_\mathrm{nl}$ does not reach the largest values observed in Fig. 3a, which implies that the spins do not fully rotate out of plane.

For a generic anisotropic spin channel, $R_\mathrm{nl}^{\beta}$ can be written as $R_\mathrm{nl}^{\beta}=\overline{R_\mathrm{nl}^{\beta}} \cos^2 (\beta-\gamma)$, where $\overline{R_\mathrm{nl}^{\beta}}$ is the non-local resistance that would be measured if the magnetization of the injector and detector were parallel to $B$. The factor $\cos^2 (\beta-\gamma)$ thus accounts for the projection of the injected spins along $B$ and the subsequent projection along the detector magnetization \cite{BR2016}. For $\zeta=1$, $\overline{R_\mathrm{nl}^{\beta}} = R_\mathrm{nl,0}$ regardless of the value of $\beta$; therefore, plotting $R_\mathrm{nl}^{\beta}$ versus $\cos^2 (\beta-\gamma)$ results in a straight line. For $\zeta \neq 1$, $R_\mathrm{nl}^{\beta}$ lies above or below the straight line depending whether $\zeta>1$ or $\zeta<1$ \cite{BR2016}. The magnitude of $R_\mathrm{nl}^{\beta}$ normalized to $R_\mathrm{nl,0}$ is shown in Fig. 4a both for the graphene/WS$_{2}$ and the reference devices (full and open symbols, respectively), as extracted from Fig. 3a and Supplementary Fig. 2. Consistent with the results in Fig. 2, $\zeta \approx 1$ for the reference, while $\zeta \gg 1$ for graphene/WS$_{2}$.

In order to find the origin of the anisotropy, the spin transport is studied as a function of $V_\mathrm{g}$, which tunes the carrier density of both graphene and WS$_2$. Figure 4b shows the obtained $\overline{R_\mathrm{nl}^{\beta}}$ vs. $\beta$ for different $V_\mathrm{g}$. Below a threshold back-gate voltage $V_\mathrm{g}^\mathrm{T}\sim -5$ V, $\overline{R_\mathrm{nl}^{\beta}}$ is nearly independent of $V_\mathrm{g}$. However, for $V_\mathrm{g} > V_\mathrm{g}^\mathrm{T}$, a rapid reduction of $\overline{R_\mathrm{nl}^{\beta}}$ is observed for all values of $\beta$. For $V_\mathrm{g} > 10$ V, no spin signal can be detected when the spins are oriented in-plane; the spin signal is recovered as soon as the out-of-plane spin component is non zero, indicating that $\zeta \gg 1$ for all values of $V_\mathrm{g}$. A vanishing $R_\mathrm{nl}$ for positive $V_\mathrm{g}$ has been recently reported in graphene/MoS$_2$ heterostructures and was attributed to the change of the MoS$_2$ channel conductivity and associated modulation of the Schottky barrier at the MoS$_2$/graphene interface \cite{yan2016,dankert2017}. As proposed in Refs. \cite{yan2016,dankert2017}, the carrier diffusion into the MoS$_2$ leads to an additional relaxation channel that suppresses the spin signal. In agreement with this interpretation, the decrease of $\overline{R_\mathrm{nl}^{\beta}}$ correlates with the increase of the WS$_2$ conductivity with $V_\mathrm{g}$. The inset of Fig. 4b shows the current $I_{\mathrm{ds}}$ vs. $V_\mathrm{g}$ when a constant driving voltage $V_{\mathrm{ds}}$ is applied between graphene and WS$_2$. It is observed that $I_{\mathrm{ds}}$ increases sharply nearby $V_\mathrm{g}^\mathrm{T}$, suggesting that only for $V_\mathrm{g} > V_\mathrm{g}^\mathrm{T}$ spins can enter WS$_2$, as proposed for MoS$_2$ (see also Supplementary Fig. 3). Since for $V_\mathrm{g} < V_\mathrm{g}^\mathrm{T}$ the carriers cannot enter WS$_2$ and $\overline{R_\mathrm{nl}^{\beta}}$ is independent of $V_\mathrm{g}$, the anisotropic spin relaxation for $V_\mathrm{g} < V_\mathrm{g}^\mathrm{T}$ must be due to proximity-induced SOC.

The spin anisotropy ratio $\zeta$ can readily be obtained from $\zeta = (\lambda_\mathrm{s}^\perp/\lambda_\mathrm{s}^\parallel)^2$. Taking $\overline{R_\mathrm{nl}^{\beta}}\sim 1.5$ $\Omega$ for $\beta \sim 90^{\circ}$ (Fig. 4b), implies that $\lambda_\mathrm{s}^\perp \sim 1$ $\mu$m, which combined with $\lambda_\mathrm{s}^\parallel \sim 0.3$ $\mu$m, results in $\zeta \approx 10$. Using these parameters, it is possible to calculate  $R_\mathrm{nl}^{\beta}$ vs. $\cos^2 (\beta-\gamma)$ (Fig. 4a, solid green line), which shows very good agreement with the experimental results considering that no adjustable parameters are used. Furthermore, $R_\mathrm{nl}^{\beta}$ vs. $B$ can be found by solving the diffusive Bloch equation \cite{BR2017}. Supplementary Figs. 4 and 5 show the calculated $R_\mathrm{nl}^{\beta}$ for a homogeneous graphene/WS$_2$ system for oblique and in-plane $B$, respectively. The general agreement between Fig. 2c and Supplementary Fig. 4a and between Fig. 3a and Supplementary Fig. 5 gives further confidence to the interpretation of our results.

Because the WS$_2$ flake is significantly narrower than the distance between F1 and F2, most of the precession occurs in graphene. This renders the precession response rather insensitive to the spin diffusion constant $D$ and the spin lifetimes within graphene/WS$_2$, as long as their product, which determine $\lambda_\mathrm{s}^\perp$ and $\lambda_\mathrm{s}^\parallel$, is kept constant (see Supplementary Information and Supplementary Fig. 6). Assuming that $D$ in graphene/WS$_2$ is of similar magnitude to that in graphene, $D\sim 0.03$ m$^2$s$^{-1}$, then $\tau_\mathrm{s}^\parallel \sim 3$ ps and $\tau_\mathrm{s}^\perp \sim 30$ ps. Here, $\tau_\mathrm{s}^\parallel$ is of the same order to that reported in graphene/MoS$_2$ with conventional spin precession measurements, $\tau_\mathrm{s} \sim 5$ ps \cite{dankert2017}. It is also very close to the values extracted from weak anti-localization experiments in graphene/WS$_2$, $\tau_\mathrm{s} \sim $ 2.5-5 ps \cite{wang2015,wang2016}.

Spin dynamics modelling and numerical simulations have been used to compute $\zeta$ in graphene interfaced with several TMDCs \cite{cummings2017}. In the case of graphene/WS$_2$ with strong intervalley scattering, $\zeta$ is calculated to lie between 20 and 200, with $\tau_\mathrm{s}^\parallel \sim 1$ ps and $\tau_\mathrm{s}^\perp \sim $ 20-200 ps. In the absence of intervalley scattering $\zeta$ decreases all the way down to 1/2, as expected for Rashba SOC, with $\tau_\mathrm{s}^\parallel \approx 2\tau_\mathrm{s}^\perp \sim $ 10 ps near the charge neutrality point. Therefore, the large $\zeta$ in our devices is not only a fingerprint of proximity-induced SOC but also indicates that intervalley scattering is important. The somewhat smaller $\zeta$ found in the experiments can originate from a number of reasons, the most straightforward being that the intervalley scattering is stronger or the SOC weaker in our devices than assumed in the model \cite{cummings2017}. The adopted parameters can also be sensitive to the number of layers in the TMDC or the specific supercells in the calculations \cite{gmitra2015,gmitra2016}. The model is developed for monolayer TMDCs; nevertheless, since the proximity effects are due to the TMDC layer that is adjacent to graphene, no dependence on the number of layers is expected, as long as the transport in graphene occurs in states within the TMDC band gap. There are other possibilities to explain the differences with the calculations, including the characteristics of the interface with graphene, and strain or twisting between the layers, which are currently not controlled in the experiments. Because the obtained $\lambda_\mathrm{s}^\perp$ is similar to the width $w$ of graphene/WS$_2$, our approach could also become insensitive for large values of $\zeta$. However, Supplementary Fig. 7 demonstrates that this limit has not been reached, in particular when $B$ is tilted slightly from the perpendicular orientation. Another aspect to be considered is the relevance of intravalley scattering and how it compares with intervalley scattering. Our measurements are all carried out at room temperature, phonon scattering can increase the  weight of intravalley  scattering and effectively reduce the anisotropy.

The spin lifetime anisotropy therefore provides insight into the physics underpinning spin and valley dynamics. Our results show that the large SOC and spin-valley coupling in a semiconducting transition metal dichalcogenide can be imprinted in graphene. In addition, they open the door for novel approaches to control spin and valley information. The out-of-plane spin component propagates through graphene/TMDC much more efficiently than the in-plane component. Thus, graphene/TMDC acts as a filter with a transmission that depends on the effective orientation of the spins that reach it and that can vary over orders of magnitude. Such a filter represents a new tool in spintronics to detect small variations in the orientation of spins arriving to it. Interfacing graphene with TMDCs can also be utilized for direct electric-field tuning of the propagation of spins and for implementing  spin-valleytronic and optospintronic \cite{gmitra2015,luo2017,avsar2017} devices in which charge, spin and valley degrees of freedom can be simultaneously used \cite{gmitra2017}, as previously proposed for TMDCs \cite{ye2016}. This is thus an important milestone for next-generation graphene-based electronics and computing.

\textit{Note}. After submission of the present manuscript, a related work studying the spin relaxation anisotropy in graphene/MoSe$_2$ has been reported \cite{ghiasi2017}. Measurements are carried out with in-plane magnetic fields at 75 K, where a spin lifetime anisotropy ratio $\zeta \approx 11$ was found.

\section{Methods}
\small{\textbf{Device Fabrication.}
Graphene/WS$_2$ heterostructures were fabricated by dry viscoelastic stamping \cite{castellanos2014}.  The experimental setup used to transfer two-dimensional crystals comprises an optical microscope with large working distance optical objectives (Nikon Eclipse Eclipse LV 100ND) and a three axis micrometer stage to accurately position the stamp.
Graphene flakes are obtained by mechanical exfoliating highly-oriented pyrolytic graphite (SPI Supplies) onto a $p$-doped Si/SiO$_2$ substrate, which is used as a back-gate. The single-layer graphene flakes are selected by optical contrast after a previous calibration with Raman measurements.  The TMDC flakes are transferred onto the stamp made of commercially available viscoelastic material (Gelpack) by exfoliation with tape, the surface of the stamp is inspected under the optical microscope to select thin and narrow flakes due to their faint contrast under normal illumination. The proximity effects are due to the adjacent layer of the TMDC to graphene. As long as the transport in the graphene is studied within the TMDC gap, no difference is expected between multilayers and monolayers. The TMDC flake on the stamp is aligned on top of the graphene target with the help of the micrometer stage, then it is pressed against the substrate and peeled off slowly. After assembling, the stacks are annealed for 1 hour at 400 $^{\circ}$C, one batch in ultra-high vacuum (10$^{-10}$ Torr) and another in high vacuum (10$^{-7}$ Torr). Annealing removes contamination between the layers as well the remaining residues from the transfer process. The two batches of samples do not show significant differences in terms of mobility and doping. The thicknesses of WS$_2$ in Devices 1 and 2 are 8 and 20 nm, respectively.
In our devices we define the contact electrodes in one (two) e-beam lithography step(s). To contact the devices all the materials were deposited by e-beam evaporation in a chamber with base pressure of 10$^{-8}$ Torr. For the one-step e-beam lithography case, we deposited  TiO$_2$/Co (1 nm/40 nm) for all the contacts. For the two-step e-beam lithography case, we deposited Ti/Pd (1 nm/30 nm) for the outer contacts (normal metal) and TiO$_2$/Co for the inner contacts (ferromagnetic metal, F1 and F2). The widths of F1 and F2 are 100 and 200 nm to ensure different coercive fields.

\textbf{Electrical characterization.}
The devices are wired to a chip carrier that is placed in a vacuum chamber. A rack-and-pinion actuator is used to change the relative angle between the chip carrier and the homogenous applied magnetic field with precision of 0.2$^{\circ}$.  Measurements were carried out at 300 K under a pressure of 10$^{-6}$ Torr. The graphene and graphene/TMDC charge transport properties are characterized by means of two- and four-terminal measurements. The contact resistance in the TiO$_2$/Co electrodes are larger than 10 k$\Omega$. The typical average electron/hole mobility is in the range of $\mu=10000$ cm$^2$V$^{-1}$s$^{-1}$ and the residual carrier density is of the order of $\sim 10^{11}$ cm$^{-2}$.}

\textbf{Data Availability.} The data that support the plots within this paper and other findings of this study are available from the corresponding authors upon reasonable request.

\vspace{5mm}

\noindent \textbf{Acknowledgments} We thank D. Torres for help in designing Fig. 1 and A. Cummings, S. Roche, J. Fabian and M. Timmermans for insightful discussions. This research was partially supported by the European Research Council under Grant Agreement No. 308023 SPINBOUND, by the European Union's Horizon 2020 research and innovation programme under grant agreement No. 696656, by the Spanish Ministry of Economy and Competitiveness, MINECO (under Contracts No MAT2016-75952-R and Severo Ochoa No. SEV-2013-0295), and by the CERCA Programme and the Secretariat for Universities and Research, Knowledge Department of the Generalitat de Catalunya 2014 SGR 56. J.F.S. acknowledges support from the MINECO Juan de la Cierva program and M.V.C. and F.B. from the MINECO Ram\'{o}n y Cajal program.

\vspace{5mm}

\noindent \textbf{Author contributions} L.A.B., J.F.S., W.S.T. and A.A. fabricated the devices and L.A.B., J.F.S. and W.S.T. performed the measurements. F.B. helped with the device fabrication and M.V.C. with the device fabrication and measurements. L.A.B. and S.O.V. analyzed the data and wrote the manuscript. All authors discussed the results and commented on the manuscript. S.O.V supervised the work.

\vspace{5mm}

\noindent \textbf{Additional Information} The authors declare no competing financial interests. Reprints and permissions information is available online at http://npg.nature.com/reprintsandpermissions. Correspondence and request for materials should be addressed to L.A.B. (antonio.benitez@icn2.cat) and S.O.V. (SOV@icrea.cat).
\newpage
\begin{figure}[ht]
\vspace{10mm}\includegraphics[width=1\linewidth]{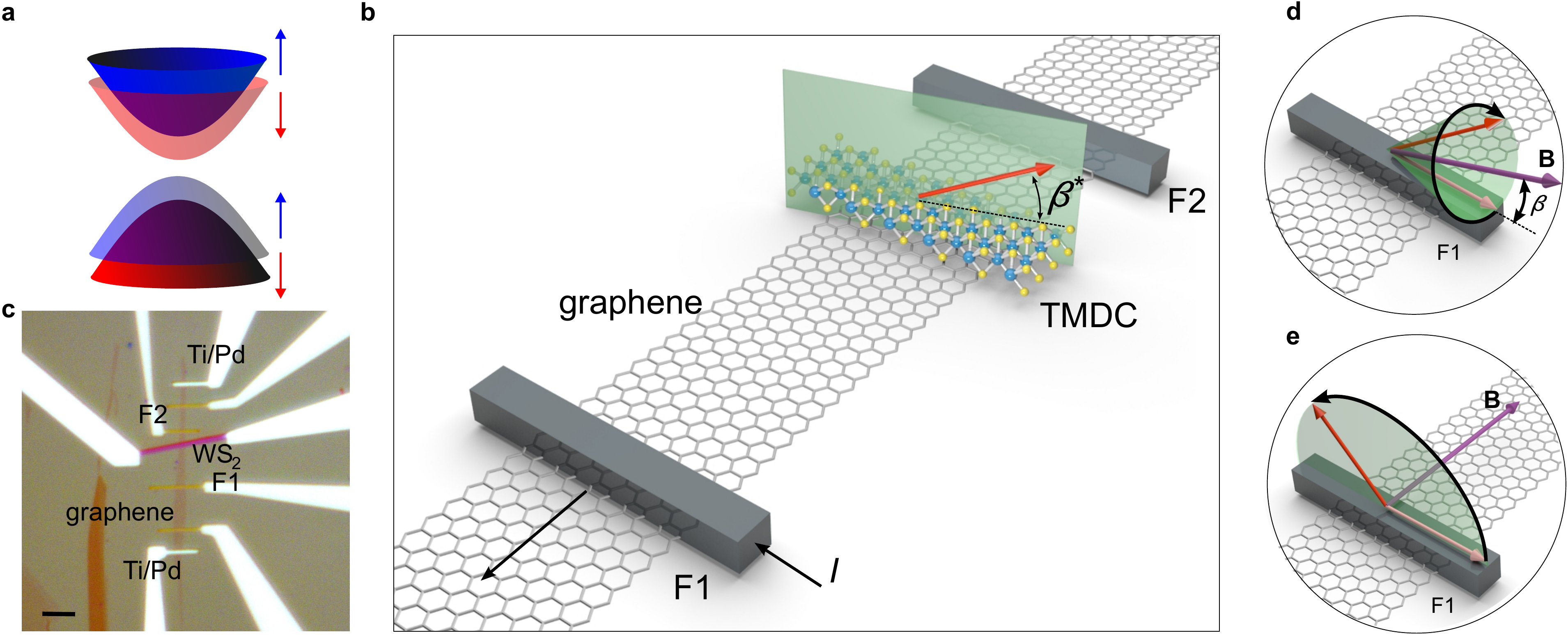}
\vspace{-2mm}
\caption{\textbf{Proximity-induced spin-orbit coupling and measurement scheme}. \textbf{a}, Representation of the graphene electronic band structure in the vicinity of the Dirac point for graphene/TMDC. The colors and arrows represent the expectation value for the out-of-plane spin component. \textbf{b}, Schematics of the device and measurement approach. The device consists of a graphene channel contacted with two ferromagnets (F1 and F2) and a transverse strip of a TMDC in between. A charge current (straight black arrows) through F1 injects spins having an orientation parallel to the F1 magnetization direction. The injected spins undergo Larmor precession under the influence of a magnetic field $B$ while diffusing towards the detector electrode (F2) (see \textbf{d} and \textbf{e}). When a spin reaches the location of the TMDC, its orientation is characterized by the angle $\beta^*$, which measures the inclination of the spin from the graphene plane. Owing to the anisotropic band structure of graphene/TMDC, the spin signal at F2 is modulated by $B$. \textbf{c}, Enhanced-contrast optical image comprising one graphene/WS$_2$ device (Device 1) and two reference graphene devices enclosing it. The bar represents 5 $\mu$m. \textbf{d}, Out-of-plane spin precession with oblique magnetic field $B$, where $B$ is applied in a plane that contains the easy axis of the ferromagnetic electrodes and that is perpendicular to the substrate. \textbf{e}, Out-of-plane spin precession with $B$ applied in-plane and perpendicular to the easy axis of F1 and F2. Spins precess in a plane perpendicular to the substrate. In \textbf{d} and \textbf{e}, the effective spin lifetime becomes sensitive to both parallel and perpendicular spin lifetimes, $\tau_{s\parallel}$ and $\tau_{s\perp}$, allowing to determine the spin relaxation anisotropy. }
\label{Fig1}
\end{figure}

\newpage
\begin{figure*}[ht!]

\vspace{-3mm}\includegraphics[width=0.7\linewidth]{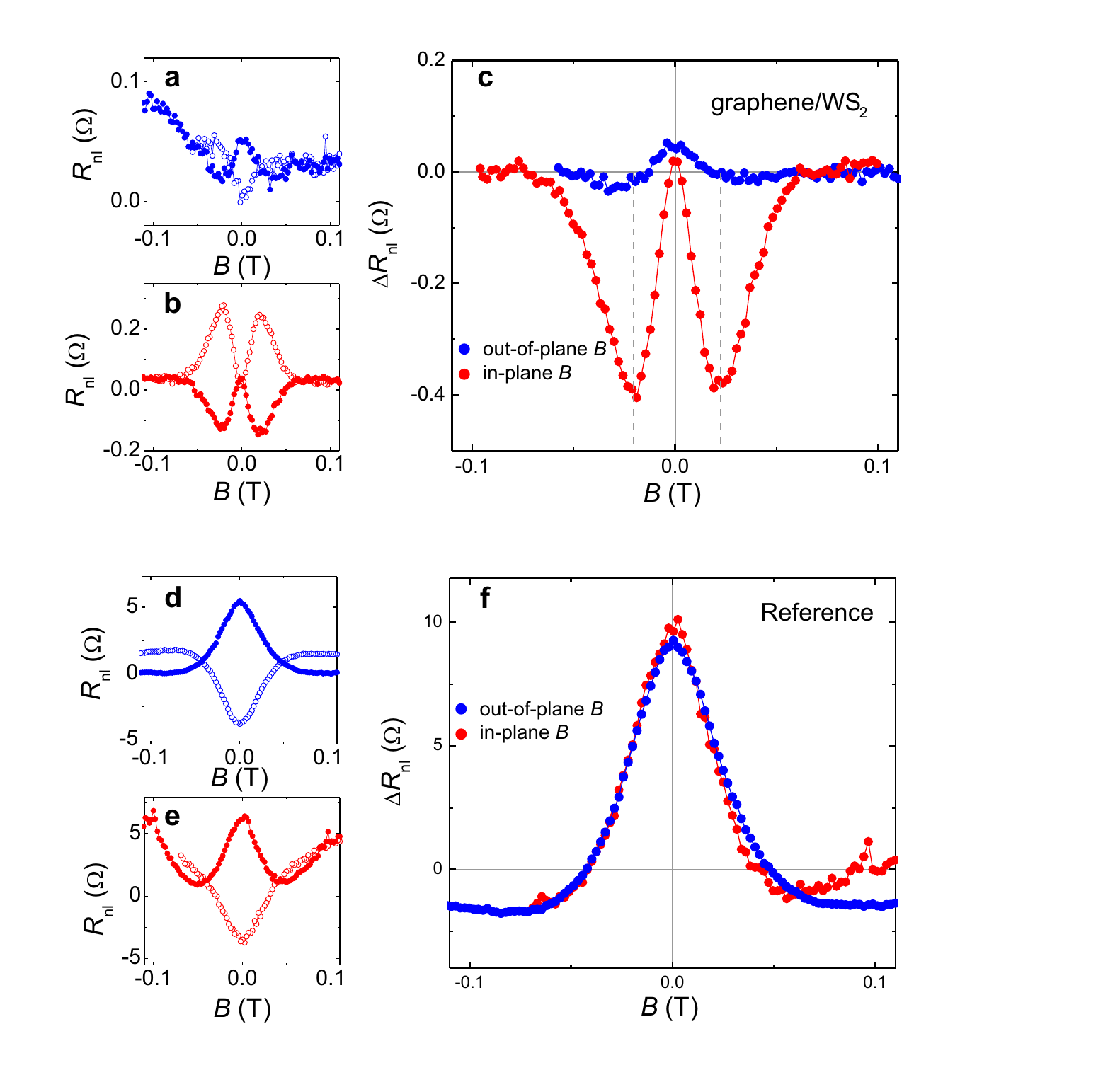}
\vspace{-8mm}
\caption{\textbf{Spin relaxation anisotropy}. Non-local resistance $R_\mathrm{nl}$ vs. $B$ for different $B$ orientations. The measurements in \textbf{a-c} are for a graphene/WS$_2$ device (Device 1) while those in \textbf{d-f} are for a reference device (without WS$_2$) in the same graphene flake. \textbf{a, d}, Standard spin precession measurements with $B$ perpendicular to the substrate. Solid (open) symbols are for parallel (antiparallel) configurations of the magnetizations of F1 and F2 ($R_\mathrm{nl}^{+}$ and $R_\mathrm{nl}^{-}$, respectively). The graphene/WS$_2$ (\textbf{a}) and reference (\textbf{d}) devices show the same qualitative response, although the magnitude of $R_\mathrm{nl}$  in the former is significantly smaller. \textbf{b, e}, Spin precession measurements with $B$ in the graphene plane as represented in Fig. 1e. The graphene/WS$_2$ (\textbf{b}) and reference (\textbf{e}) devices display different behaviour. $R_\mathrm{nl}$ in the reference device does not vary with $B$ orientation, neither in magnitude nor in the precession features. In the graphene/WS$_2$ device $R_\mathrm{nl}$ increases in magnitude when spins rotate out of the graphene plane, rapidly changing sign; $R_\mathrm{nl}$ presents extrema at $B \ \sim \pm 20$ mT with a magnitude that exceeds $|R_\mathrm{nl}(B=0)|$. \textbf{c, f},  $\Delta R_\mathrm{nl}= R_\mathrm{nl}^{+}-R_\mathrm{nl}^{-}$ versus $B$ as extracted from \textbf{a, b} and \textbf{d, e}. The blue (red) plots are for perpendicular (in-plane) $B$. The dashed lines in \textbf{c} mark the position of the extrema for the measurements with in-plane $B$.} \label{Fig2}
\end{figure*}

\newpage
\begin{figure*}[ht!]
\includegraphics[width=0.6\linewidth]{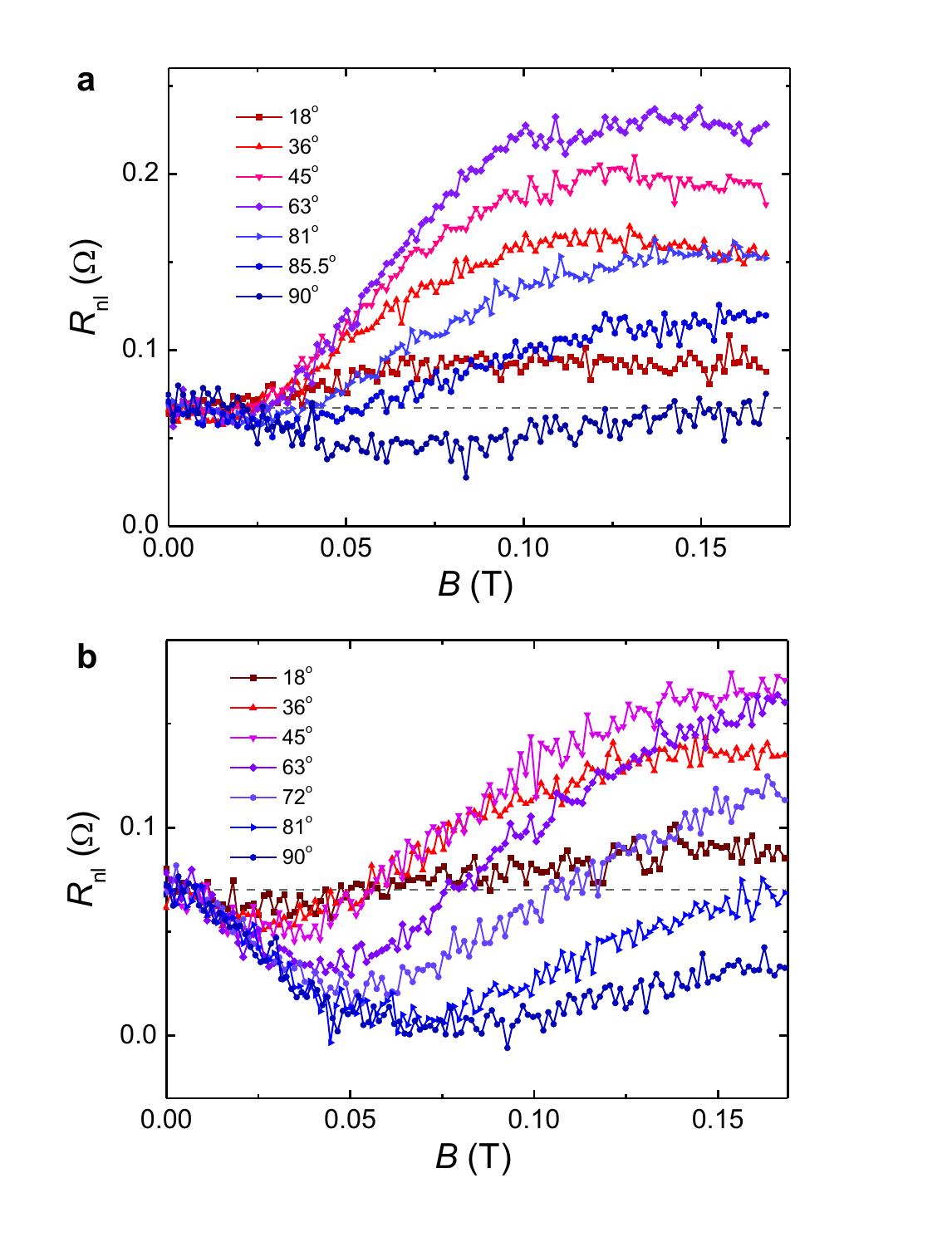}
\caption{\textbf{Spin precession measurements under oblique magnetic fields}. \textbf{a,} Representative subset of experimental spin precession curves for different $\beta$ as a function of $B$, when F1 is used as spin injector. The precession data are acquired after preparing a parallel magnetization configuration of F1 and F2 using Device 2 at $V_\mathrm{g}=-15$ V. Dephasing of the precessing component is observed at $B$ larger than $\sim 0.12$ T. The horizontal dashed line is the non-local resistance at $B=0$, $R_\mathrm{nl,0}$, which coincides with $R_\mathrm{nl}$ at $\beta=0^{\circ}$ in the parallel configuration. Similarly to Fig. 2b, $R_\mathrm{nl}$ surpasses $R_\mathrm{nl,0}$ as soon as the spins rotate out-of-plane. \textbf{b}, Representative subset of experimental spin precession curves for different $\beta$ as a function of $B$ when F2 is used as spin injector. Because graphene/WS$_2$ is close to the injector, dephasing of the precessing component is not achieved at its location and $R_\mathrm{nl}$ becomes dependent of $B$ in the full $B$ range. The distance between F1 and F2 is $11$ $\mu$m , and between F1 and WS$_2$ is $7.5$ $\mu$m.} \label{Fig3}
\end{figure*}

\newpage
\begin{figure*}[ht!]
\includegraphics[width=0.6\linewidth]{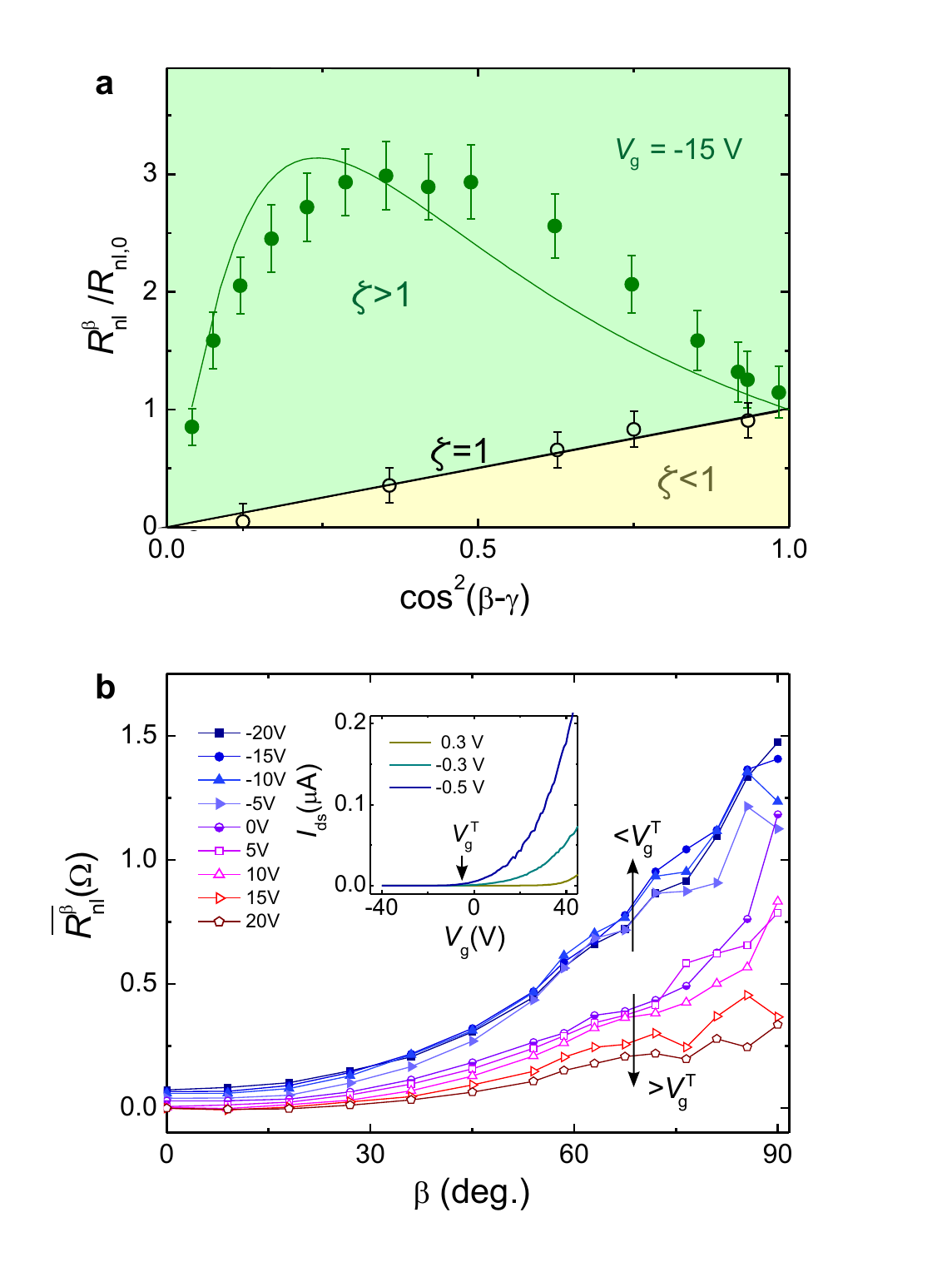}
\caption{\textbf{Spin lifetime anisotropy ratio, $\zeta$}. \textbf{a}, $R_\mathrm{nl}^{\beta}$ normalized by $R_\mathrm{nl,0}$ as a function of $\cos^2(\beta-\gamma)$, with $\gamma=\gamma(\beta,B)$. The data represented by solid symbols are extracted from Fig. \ref{Fig3}a at $B= 0.16$ T. The solid green line represents the modeled response for $\zeta=10$.  The data represented by open symbols are extracted from Supplementary Fig. 2 and correspond to a reference device. The error bars reflect the propagation of the uncertainties in $R_{nl}$ and $R_{nl,0}$ deriving from the measurement noise in Fig. \ref{Fig3}a and Supplementary Fig. 2. In this case, $\zeta\approx 1$, as shown by the straight black line. \textbf{b}, $\overline{R_\mathrm{nl}^{\beta}}$ as a function of $\beta$ for the indicated back-gate voltages $V_\mathrm{g}$. For $V_\mathrm{g} < V_\mathrm{g}^\mathrm{T} \approx -5$ V, $\overline{R_\mathrm{nl}^{\beta}}$ is independent of $V_\mathrm{g}$ but decreases rapidly for $V_\mathrm{g} > V_\mathrm{g}^\mathrm{T}$. Inset, transfer characteristics $I_{\mathrm{ds}}$ versus $V_\mathrm{g}$ for different bias voltage $V_{\mathrm{ds}}$ in graphene/WS$_2$; $V_\mathrm{g}^\mathrm{T}$ coincides with the back-gate voltage at which $I_{\mathrm{ds}}$ is observed.} \label{Fig4}
\end{figure*}
\end{document}